\RequirePackage[2018-12-01]{latexrelease}
\documentclass[showpacs,showkeys,11pt,
preprint,preprintnumbers,nofootinbib,
groupedaddress,superscriptaddress,amsmath,amssymb]{revtex4}
\usepackage{cancel}
\usepackage{xcolor}
\usepackage{tabularx}
\usepackage{amsfonts}
\usepackage{amsmath}
\usepackage{amsfonts}
\usepackage{graphicx,subfigure}
\usepackage{caption}
\usepackage{subcaption}
\usepackage[export]{adjustbox}
\usepackage{verbatim} 
\usepackage{epsfig}
\usepackage{url}
\usepackage{float}
\usepackage{multirow}
\usepackage{hhline}
\usepackage{feynmp}
\usepackage{booktabs}
\usepackage{csquotes}

\newcommand {\be}{\begin{equation}}
\newcommand {\ee}{\end{equation}}
\newcommand {\ba}{\begin{eqnarray}}
\newcommand {\ea}{\end{eqnarray}}

\begin{document}
\title{Unraveling Dirac Magnetic Monopoles with Muon Beams at TeV Energies Using Machine Learning}


\author{M. Tayyab Javaid}
\affiliation{Federal Urdu University of Arts, Science and Technology, Islamabad, Pakistan}

\author{Haroon Sagheer}
\affiliation{Riphah International University, Islamabad}

\author{M. Danial Farooq}
\affiliation{Federal Urdu University of Arts, Science and Technology, Islamabad, Pakistan}

\author{Ijaz Ahmed \footnote{Corresponding author: }}
\email{ijaz.ahmed@fuuast.edu.pk}
\affiliation{Federal Urdu University of Arts, Science and Technology, Islamabad, Pakistan}

\author{Jamil Muhammad}
\email{mjamil@konkuk.ac.kr}
\affiliation{Department of Physics, Konkuk University, Seoul 05029, South Korea}

\date{\today}

\begin{abstract}
The focus of this paper is the production of magnetic monopoles via the Drell-Yan and photon-fusion mechanisms to generate velocity-dependent scalar, fermionic, and vector monopoles of spin angular momentum 0, 1/2, 1, respectively at a future muon collider. A computational study compares the monopole pair-production cross-sections for both methods at various center-of-mass energies ($\sqrt{s}$) with different magnetic dipole moments. The comparison of kinematic distributions of monopoles at the generator and reconstructed level is demonstrated for both DY and PF mechanisms. We demonstrate the observability of magnetic monopoles against the most relevant Standard Model background using multivariate analysis techniques. Specifically, we apply three different classifiers based on neural networks, e.g., Boosted Decision Trees, Multilayer Perceptrons, and Likelihood methods, to evaluate their effectiveness. Our results highlight the efficiency and robustness of these approaches in distinguishing magnetic monopole signals from background noise. Quantitatively, our analysis shows that the Photon-Fusion mechanism significantly dominates at high energies; for a spin-1/2 monopole with a mass of 1.5 TeV at $\sqrt{s}=30$ TeV, the PF cross-section (13.33 Pb) is approximately 30 times larger than the Drell-Yan process (0.44 Pb). Furthermore, the Boosted Decision Tree (BDT) classifier outperforms other methods, reaching a signal efficiency of 0.97 and a statistical significance of 22.80 at an optimal cut of 0.042, with an AUC value of 0.989, underscoring the high discovery potential for magnetic monopoles in multi-TeV muon collisions.
\end{abstract}

\maketitle

\section{Introduction}
Ever since Dirac's groundbreaking work in 1931, physicists have been fascinated by magnetic monopoles, which are particles believed to carry a net magnetic charge. Dirac showed that the quantization of electric charge could be explained by the presence of a single magnetic monopole. This concept is expressed in the elegant relation $eg=n\hbar c/2$, where e stands for electric charge, g for magnetic charge, $\hbar$ for reduced Planck constant, c for speed of light, and n for an integer. As of yet, magnetic monopoles have no direct scientific proof, despite their theoretical appeal and potential to unite fundamental forces \cite{MMDirac}. They continue to be one of physics' biggest mysteries.\\
To find these mysterious particles, multiple experimental attempts have been made throughout the years. More recent searches at the Large Hadron Collider (LHC) and the Tevatron have pushed the bounds farther, while earlier experiments, including those at SLAC and LEP, looked for monopoles in $e^+ e^-$ collisions \cite{MMExp}. The ATLAS collaboration, for instance, has established strict constraints on monopole production at $\sqrt{s}$=13 TeV at the LHC, recently updated with higher luminosity data \cite{Aad}, excluding monopoles below specific weights based on their theoretical characteristics \cite{ATLAS}. Similar to this, specialized detectors such as MoEDAL (Monopole and Exotics Detector at the LHC) have made it possible to record the distinctive, highly ionizing traces that monopoles may leave \cite{MOEDAL1}. Along with colliders, astrophysical studies, such as magnetic flux measurements, and cosmic ray data analysis, have also imposed constraints on monopole features, frequently reaching mass scales that are not yet reachable by colliders \cite{Gamberg}.\\
In the pursuit of magnetic monopoles, muon beams with TeV energy present a novel new angle. Because muons involve far smaller radiation losses during acceleration than protons or electrons, high-energy, high-intensity beams that are perfect for rare particle searches can be produced \cite{MuonCollider}, providing a unique 'muon smasher' environment for new physics  \cite{Fabbrichesi, Long}. The synthesis of monopoles, whose interaction cross-section, monopole, is sensitive to the magnetic charge g, may be accomplished effectively through processes such as Drell-Yan production or photon-photon fusion in muon collisions, according to theoretical models. The production rate, for example, grows with $g^4$ in photon-fusion mechanisms, indicating the ability of muon beams to study this special characteristic. However, given the possibly new and high-ionization signatures monopoles would leave behind, how to separate monopole signals from an overwhelming set of conventional model events is a real challenge.
Physicists now approach these problems in a different way thanks to the development of machine learning (ML). In order to improve event selection and signal identification, researchers can train algorithms on simulated monopole events. Convolutional neural networks (CNNs), for instance, excel at deciphering complex detector data, but other methods, like ensemble learning and decision trees, can identify uncommon occurrences hidden in the background. To improve sensitivity, data-driven methods have already started to be incorporated into initiatives like those at MoEDAL \cite{MOEDAL2}. Furthermore, recent research from the CMS and ATLAS collaborations has shown how machine learning (ML) may greatly increase detection efficiency, making it a vital tool for contemporary particle physics \cite{ATLAS1, CMS1, CMS2, CMS3, Qu}, especially when utilizing deep learning for signal discrimination \cite{Guest, Baldi}.

It should be noted that the large magnitude of the Dirac magnetic charge ($g_D \approx 68.5e$) implies an extremely strong coupling, which places the theory in a non-perturbative regime where standard Feynman diagrammatic expansions are not formally under control. Nevertheless, following the established conventions of the ATLAS and MoEDAL collaborations \cite{ATLAS, MOEDAL1}, we utilize tree-level cross-section estimates as essential phenomenological benchmarks. These benchmarks are necessary to provide a quantitative framework for evaluating experimental sensitivity and developing signal-background discrimination strategies.
\section{Dirac Model Description}
In 1931, Paul A. M. Dirac proposed a groundbreaking theoretical framework introducing the concept of the \textit{magnetic monopole} - a hypothetical particle that serves as the magnetic analogue of an electric point charge \cite{MMDirac}. Although no experimental evidence for such a particle has yet been found [3-5], despite extensive searches across various collider and non-collider experiments \cite{Lu, Vidal}. Dirac’s model remains one of the most elegant theoretical extensions of classical and quantum electrodynamics. It provides a natural explanation for the quantisation of electric charge and restores complete symmetry to Maxwell’s equations.

\subsection{Dirac Monopoles}

Dirac’s investigation focused on whether the established theories of electromagnetism and quantum mechanics could accommodate the existence of isolated magnetic charges. He demonstrated that the inclusion of a magnetic monopole in these frameworks implies that the product of the electric charge $q$ and the magnetic charge $g$ must be quantised according to the condition:
\begin{equation}
    qg = n\frac{\hbar}{2},
\end{equation}
where $n$ is an integer. This result, known as the \textit{Dirac quantisation condition}, reveals that the mere existence of a single magnetic monopole anywhere in the universe would suffice to explain why all observable electric charges are quantised in integer multiples of a fundamental unit. Thus, the Dirac monopole elegantly connects quantum mechanics and the discrete nature of electric charge.

\subsection{The Dirac String}

When defining the magnetic vector potential $\vec{A}$ for a monopole located at the origin, one encounters an unavoidable singularity. To preserve gauge invariance, Dirac introduced the concept of an invisible, infinitesimally thin line of singularity known as the \textit{Dirac string}, which carries magnetic flux from infinity to the monopole. The magnetic field of a Dirac monopole can be expressed as:
\begin{equation}
    \vec{B}_D(\vec{r}) = \nabla \times \vec{A}_D = \frac{\mu_0}{4\pi r^3}|\vec{\mu}_D|(2\cos\theta\,\hat{r} + \sin\theta\,\hat{\theta}),
\end{equation}
where $\vec{\mu}_D$ is the magnetic moment and $\theta$ the polar angle. The Dirac string is a mathematical artifact—it can be positioned arbitrarily and made physically unobservable if the phase of an electron wavefunction encircling it changes by an integer multiple of $2\pi$. This requirement ensures the string’s invisibility and enforces charge quantisation.
\subsection{Charge Quantization Condition}
In quantum electrodynamics, the potentials $\phi$ and $\vec{A}$ acquire direct physical significance through phenomena such as the \textit{Aharonov–Bohm effect}. For a charged particle moving around the Dirac string, the phase shift of its wavefunction is given by:
\begin{equation}
    \Delta \phi = \frac{q}{\hbar}\oint \vec{A}\cdot d\vec{l} = \frac{\mu_0 q g}{\hbar}.
\end{equation}
To ensure the wavefunction remains single-valued, this phase shift must be an integer multiple of $2\pi$, leading to the quantisation condition:
\begin{equation}
    qg = n\frac{h}{\mu_0}.
\end{equation}
Alternatively, by imposing angular momentum quantisation on a system consisting of a monopole and an electric charge, the same relation is obtained, reinforcing the universality of the result.

From this, the \textit{Dirac magnetic charge} $g_D$ is defined as:
\begin{equation}
    g_D = \frac{ce}{2\alpha}\approx 68.5 e.
\end{equation}
where $e$ is the elementary electric charge and $\alpha$ the fine-structure constant. Numerically, $g_D \approx 68.5\,e$, indicating that a magnetic monopole would possess an extremely strong electromagnetic coupling $(\alpha_m \approx 34)$, placing them in the non-perturbative regime of quantum field theory \cite{Baines}.

\section{Collider Phenomenology of Magnetic Monopole} 
We analyze monopole production at a future Muon Collider, which offers high center-of-mass energies ($\sqrt{s}$) and clean collision environments \cite{MuonCollider, Stratakis}, which are essential for precision physics at the multi-TeV scale \cite{Bals, Ali}. We utilize \textbf{MadGraph5} \cite{Alwall} to simulate production via Drell-Yan (DY) and Photon-Fusion (PF) mechanisms, incorporating a $\beta$ dependent coupling $g({\beta})=\textit{g}_{D}{\beta}$ 
to regularize the cross-section at the threshold \cite{Baines}. The simulation incorporates electroweak parton distribution functions (PDFs) to accurately model the photon content of the high-energy muon beams \cite{Han}.

Given that the magnetic coupling is of order $1/\alpha_{em}$, higher-order photon contributions are not suppressed, and the reliability of the one-photon exchange approximation remains a subject of ongoing theoretical discussion in the non-perturbative regime. In this study, we treat the production cross-sections derived from the Drell-Yan and photon-fusion mechanisms as effective baseline estimates. While these values carry significant theoretical uncertainties due to the large coupling constant, they serve as the standard phenomenological reference models used by current experiments to explore the kinematic reach of future colliders \cite{ATLAS, MOEDAL1, Baines}. Furthermore, the primary focus of this work is the relative difference in kinematic distributions between various spin hypotheses and the optimization of multivariate classifiers, which are primarily sensitive to the event topology and kinematic shapes rather than the absolute normalization of the cross-section.

\begin{figure}[H]
  \centering
    \includegraphics[width=6cm,height=4cm]{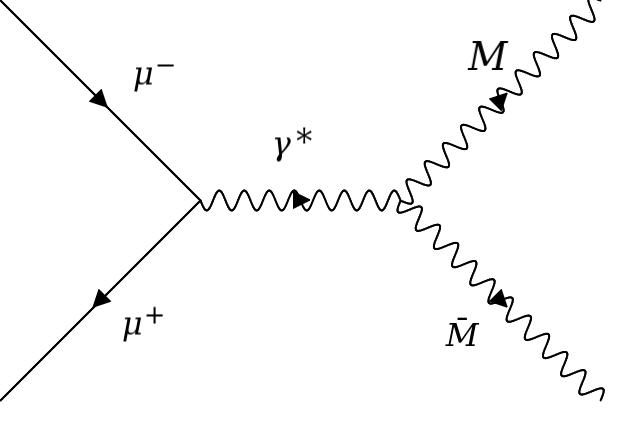}
    \hspace{2cm}
    \includegraphics[width=4.5cm,height=5cm]{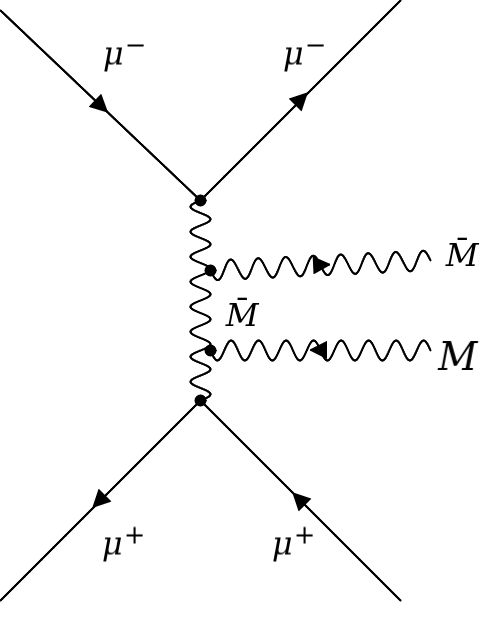}
  \caption{(Left) Monopole pair production via Drell-Yan and (Right) via Photon Fusion.}
  \label{fig:4.5}
\end{figure}
 \subsection{Spin 0 Magnetic Monopole}
\begin{figure}[H]
  \centering
  \begin{minipage}{0.470\textwidth}
    \includegraphics[width=\linewidth]{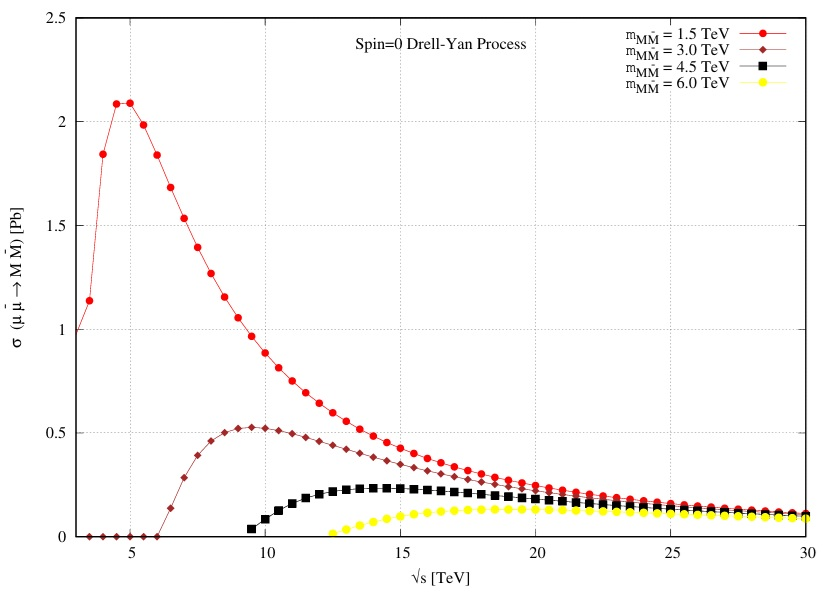}
  \end{minipage}
  \hfill
  \begin{minipage}{0.470\textwidth}
    \includegraphics[width=\linewidth]{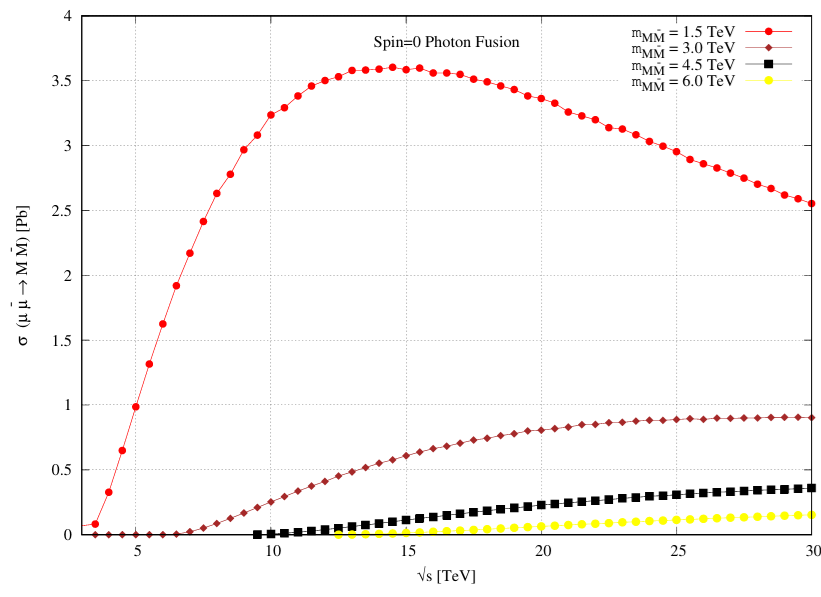}
  \end{minipage}
  \caption{Total cross section at $\sqrt{s}$ =1.5 TeV - 30 TeV versus monopole mass for spin-0 via DY and PF.}
  \label{fig:4.5}
\end{figure}
Figure 1 delineates the total production cross-section for spin-0 magnetic monopoles as a function of the center-of-mass energy ($\sqrt{s}$), ranging from 1.5 TeV to 30 TeV. The analysis compares two distinct production mechanisms: the Drell-Yan process (left panel) and Photon Fusion (right panel).
Distinct kinematic behaviors are observed for the different monopole mass hypotheses 
$(m_{MM} =$ 1.5, 3.0, 4.5, and 6.0 TeV). As expected, the production rate exhibits a steep dependence on the monopole mass; lighter candidates (1.5 TeV) show significantly higher cross-sections, whereas heavier states are kinematically suppressed and require higher collision energies to become accessible. Particularly, the right panel shows a steep rise initially. For heavy monopoles, the probability drops precipitously as mass increases. However, at ($\sqrt{s}$) = 30 TeV,  the PF mechanism maintains a significant rate due to the enhanced photon luminosity at high energies \cite{Franceschini}.
Both of these plots reveal that the probability of generating these heavy particles drops precipitously as the monopole mass increases.
\subsection{Spin 1/2 Magnetic Monopole}
\begin{figure}[H]
  \centering
  \begin{minipage}{0.45\textwidth}
    \includegraphics[width=\linewidth]{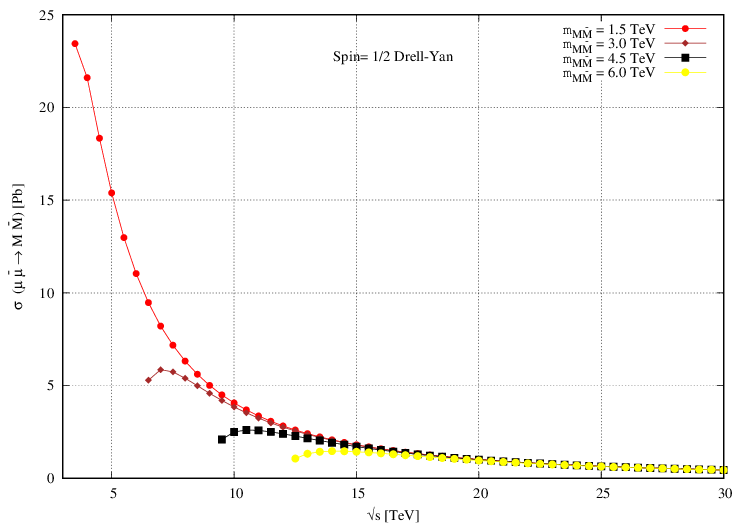}
  \end{minipage}
  \hfill
  \begin{minipage}{0.45\textwidth}
    \includegraphics[width=\linewidth]{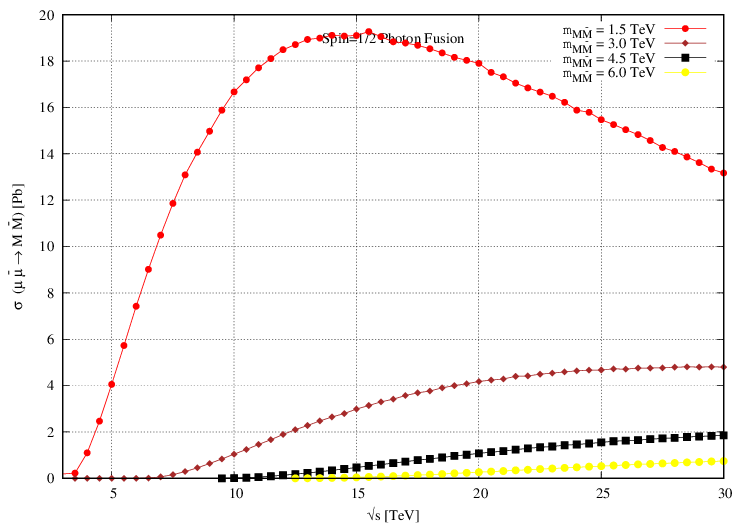}
  \end{minipage}
  \caption{Total cross section at $\sqrt{s}$ =1.5 TeV - 30 TeV versus monopole mass for spin-1/2 via DY and PF.}
  \label{fig:4.5}
\end{figure}
Figure 2 presents the calculated total cross-sections for the production of fermionic (spin-1/2) magnetic monopoles. The variation of the cross-section is plotted against the center-of-mass energy ($\sqrt{s}$), for the Drell-Yan (left panel) and Photon Fusion (right panel) channels.
A comparison with the scalar scenario reveals a significant enhancement in production rates for the spin-1/2 case. In the Drell-Yan process, the cross-section for a 1.5 TeV monopole reaches values exceeding 20 Pb at lower energies, driven by the additional spin-dependent coupling terms in the scattering amplitude. The curve exhibits a sharp inverse dependence on ($\sqrt{s}$), for lighter masses, rapidly decreasing as the energy scale expands. This behavior aligns with recent theoretical predictions for high-energy lepton colliders \cite{Dougall}.
In contrast, the Photon Fusion mechanism displays a resonant-like structure, particularly evident for the 1.5 TeV mass hypothesis, where the cross-section peaks near 
$\sqrt{s} = 13$ TeV before gradually diminishing. For heavier mass points (4.5 TeV and 6.0 TeV), the onset of production is kinematically delayed, requiring significantly higher collision energies to overcome the phase-space suppression."

\subsubsection{$\beta$ Dependent spin=1/2}
The production of spin–$\frac{1}{2}$ magnetic monopoles at high-energy colliders depends strongly on the particle velocity $\beta = v/c$, since the magnetic coupling is $\beta$-dependent.  According to Dirac's quantization, the fundamental magnetic charge is $g_D = \frac{e}{2\alpha} \approx 68.5e$. However, for phenomenological consistency at a collider, we adopt a velocity-dependent effective coupling:
\begin{equation}
    g(\beta) = g_D \beta^n
\end{equation}
where $n=1$ in this study. This $\beta$-dependence is a standard regularization choice implemented in recent monopole studies [13, 17, 21]. Physically, this ensures that the interaction strength vanishes at the production threshold ($\beta \to 0$), which is required to maintain Lorentz invariance in the cross-section and to avoid unphysical divergences at low velocities.
The interaction Lagrangian for a spin–$\frac{1}{2}$ monopole can be expressed as:
\begin{equation}
\mathcal{L} = \bar{\psi}(i\gamma^{\mu}\partial_{\mu} - m)\psi - \frac{1}{2}g(\beta)\bar{\psi}\sigma^{\mu\nu}F_{\mu\nu}\psi,
\end{equation}
where the $\beta$-dependent term governs the monopole–photon interaction strength.
The interaction Lagrangian in Eq.~(7) utilizes a magnetic "dipole-type" coupling ($\sigma^{\mu\nu} F_{\mu\nu}$). While a fundamental monopole is defined by its magnetic charge, for particles with non-zero spin, the derivative coupling represented by the dipole form is the standard effective field theory (EFT) approach for investigating monopole phenomenology at high-energy colliders [13, 19]. This specific Lagrangian structure is chosen for two reasons: first, it allows for the inclusion of an anomalous magnetic moment ($\kappa$), and second, it provides a consistent framework for comparing scalar, fermionic, and vector monopoles using the same MadGraph5\_aMC@NLO implementation used by the MoEDAL and ATLAS collaborations. By using this effective form, we provide a benchmark for how the particle's spin-dependent interactions influence the observed kinematic distributions and machine learning discrimination.

In photon–fusion processes,
\begin{equation}
\gamma\gamma \rightarrow M\bar{M},
\end{equation}
the production cross-section behaves as 
\begin{equation}
\sigma_{\gamma\gamma} \propto \frac{(g(\beta))^{4}}{s}\,\beta\left(1 - \frac{\beta^2}{3}\right),
\end{equation}
while for the Drell–Yan mechanism,
\begin{equation}
\mu\bar{\mu} \rightarrow \gamma^* \rightarrow M\bar{M},
\end{equation}
it follows
\begin{equation}
\sigma_{DY} \propto \frac{(g(\beta))^2}{s}\,\beta(3 - \beta^2).
\end{equation}
These relations imply $\sigma_{DY}\!\sim\!\beta^3$ and $\sigma_{\gamma\gamma}\!\sim\!\beta^5$ \cite{Farrar}, indicating that Drell–Yan dominates near threshold, whereas photon fusion becomes the leading process at higher energies ($\sqrt{s} \gtrsim 13$~TeV). The $\beta$-dependent formulation therefore provides a consistent framework for describing spin–$\frac{1}{2}$ monopole production and aligns well with collider observations from ATLAS and MoEDAL, which constrain monopole masses in the 1–6~TeV range.
\begin{table}[h!]
\caption{Cross sections at $\sqrt{s_{\mu\bar{\mu}}}=30~\text{TeV}$ for spin-1/2 monopoles with $\beta$-dependent coupling of Photon Fusion and Drell-Yan process at various values of the $\kappa$ parameter.}
\begin{center}
 \begin{tabular}{|c|c|c|c|c|c|}
\hline
\textbf{Monopole mass} & $\boldsymbol{\beta}$ &
{$\kappa=0$} & {$\kappa=10$} & {$\kappa=100$} & {$\kappa=10^{4}$} \\
\hline
\textbf{(TeV)} & \multicolumn{5}{|c|}{$\gamma \gamma \rightarrow M \bar{M}$ } \\   \hline
 1.5 & 0.9515 & 13.33 & $2.44\times10^{20}$ & $2.44\times 10^{24}$ & $2.47\times10^{32}$ \\
3.0 & 0.8871 & 4.78 & $1.90\times10^{20}$ & $1.91\times10^{24}$ & $1.94\times10^{32}$ \\
4.5 & 0.7882 & 1.79 & $1.41\times10^{20}$ & $1.40\times10^{24}$ & $1.40\times10^{32}$ \\
6.0 & 0.6390 & 0.69 & $9.33\times10^{19}$ & $9.29\times10^{23}$ & $9.28\times10^{31}$ \\
\hline
& \multicolumn{5}{|c|}{$\mu \bar{\mu} \rightarrow M \bar{M}$ } \\   \hline
1.5 & 0.9949 & 0.44 & $2.04\times10^{10}$ & $2.04\times10^{12}$ & $2.04\times10^{16}$ \\
3.0 & 0.9798 & 0.43 & $2.07\times10^{10}$ & $2.07\times10^{12}$ & $2.07\times10^{16}$ \\
4.5 & 0.9539 & 0.41 & $2.09\times10^{10}$ & $2.09\times10^{12}$ & $2.09\times10^{16}$ \\
6.0 & 0.9165 & 0.37 & $2.07\times10^{10}$ & $2.07\times10^{12}$ & $2.07\times10^{16}$ \\
 \hline 
\end{tabular}%
\label{table:5}
\end{center}
\end{table}
A comparative analysis of the competing production mechanisms is provided in Table I. The data contrasts the yield of the Photon Fusion $(\gamma \gamma \rightarrow MM)$
 process against the direct muon annihilation ({$\mu \bar{\mu} \rightarrow M \bar{M}$ })  channel across a range of monopole masses.
The numerical results confirm the absolute dominance of the photon-induced channel. For a benchmark mass of 1.5 TeV with standard coupling $(\kappa =0)$
, the Photon Fusion cross-section (13.33 Pb) is approximately 30 times larger than that of the Drell-Yan process (0.44 Pb).
This disparity becomes even more pronounced in the presence of anomalous couplings. For $(\kappa =10)$, the Photon Fusion rate jumps to the order of 
$10^{20}$, whereas the Drell-Yan rate remains in the $10^{10}$
 regime. This observation suggests that for spin-1/2 monopoles, the photon fusion mechanism serves as the primary discovery channel, offering a sensitivity reach that far exceeds that of direct lepton annihilation.

\subsection{Spin 1 Magnetic Monopole}
\begin{figure}[H]
  \centering
  \begin{minipage}{0.45\textwidth}
    \includegraphics[width=\linewidth]{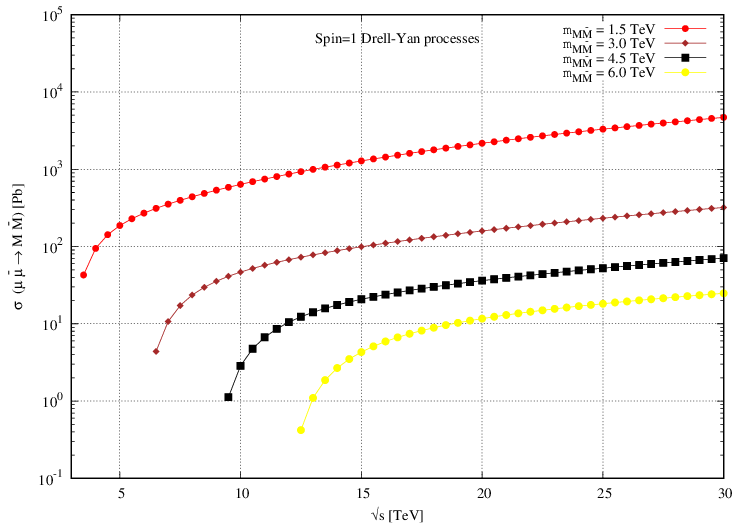}
  \end{minipage}
  \hfill
  \begin{minipage}{0.45\textwidth}
    \includegraphics[width=\linewidth]{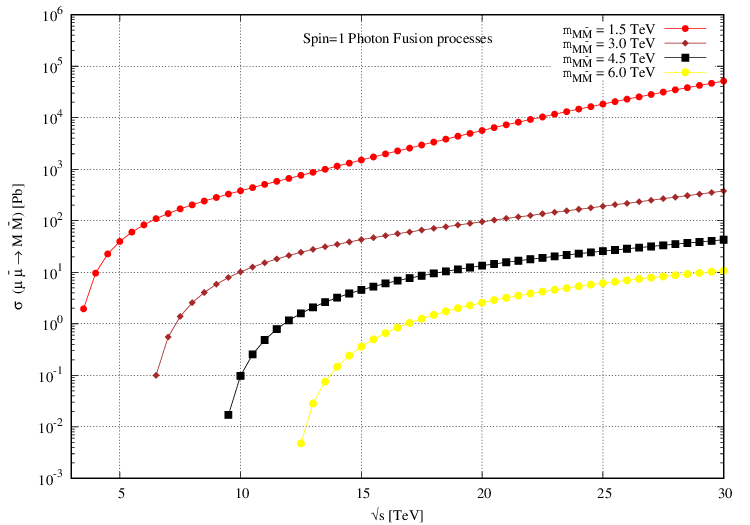}
  \end{minipage}
  \caption{Total cross section at $\sqrt{s}$ =1.5 TeV - 30 TeV versus monopole mass for spin-1 via DY and PF.}
  \label{fig:4.5}
\end{figure}
Figure 3 depicts the total production cross-sections for vector (spin-1) magnetic monopoles, utilizing a logarithmic scale to accommodate the expansive range of the predicted values. The left and right panels correspond to the Drell-Yan and Photon Fusion mechanisms, respectively, across a center-of-mass energy span of 1.5 to 30 TeV.
A striking feature of the vector hypothesis is the magnitude of the production rates. Unlike the scalar and fermionic models, the spin-1 cross-sections do not exhibit a turnover or decrease at high energies; rather, they demonstrate a monotonic rise with energy. This violates unitarity at very high energies if not cut off, a known feature of effective theories for massive vector bosons \cite{Khoze}.
For the lightest mass case ($m_{MM} =$ 1.5 TeV), the Photon Fusion cross-section climbs rapidly, exceeding $10^{5}$ Pb at the upper energy limit. This behavior is indicative of the energy-dependent growth characteristic of massive vector boson scattering in effective field theories, suggesting that if monopoles possess vector quantum numbers, their discovery signal would be overwhelmingly dominant over background processes.

\subsubsection{$\beta$ Dependent spin1}
 spin–$1$ (vector) magnetic monopoles, the same $\beta$-dependent coupling defined previously applies, ensuring finite interaction strength at low velocities. The interaction can be formulated using a Proca-type Lagrangian,
\begin{equation}
\mathcal{L} = -\frac{1}{2}(D_\mu W_\nu - D_\nu W_\mu)^\dagger (D^\mu W^\nu - D^\nu W^\mu) + m^2 W_\mu^\dagger W^\mu - i g(\beta) F^{\mu\nu} W_\mu^\dagger W_\nu,
\end{equation}
where $W_\mu$ denotes the vector monopole field. The $\beta$-dependence again regulates the coupling near threshold and enhances it at relativistic energies. Compared to the spin–$\frac{1}{2}$ case, the photon–fusion cross-section rises more steeply with $\beta$, while Drell–Yan remains subdominant. The leading-order dependencies can be summarised as $\sigma_{\gamma\gamma}\!\sim\!\beta^5$ and $\sigma_{DY}\!\sim\!\beta^3$, with photon fusion becoming the dominant production mode for monopole masses above $\sim 1.5$--$2~\text{TeV}$ at LHC energies. The higher magnetic moment of the spin–$1$ monopole further enhances the cross-section, making such states particularly favourable in photon-rich environments. These $\beta$-dependent effects align with theoretical expectations and collider constraints reported by ATLAS and MoEDAL 
Cof Photon-fusion and Drell-Yan production at 
\begin{table}[ht]
\centering
\begin{tabular}{|c|c|c|c|c|c|}
\hline
\textbf{Monopole mass} & \textbf{$\beta$} & \(\kappa = 0\) & \(\kappa = 10\) & \(\kappa = 100\) & \(\kappa = 10^4\) \\
\textbf{[TeV]} & & & & & \\
\hline
1.5 & 0.9515 & \(10^7\) & \(7.16 \times 10^8\) & \(7.41 \times 10^{12}\) & \(7.43 \times 10^{20}\) \\
3.0 & 0.8871 & 44.83 & \(3.36 \times 10^6\) & \(3.55 \times 10^{10}\) & \(3.35 \times 10^{18}\) \\
4.5 & 0.7882 & 3.56 & \(1.07 \times 10^5\) & \(1.11 \times 10^9\) & \(1.11 \times 10^{17}\) \\
6.0 & 0.6390 & 0.07 & 1453 & \(1.47 \times 10^7\) & \(1.47 \times 10^{15}\) \\
\hline
\end{tabular}
\caption{Monopole production cross-section in pb at various masses in TeV and magnetic moments}
\end{table}
Table II enumerates the production cross-sections for vector (spin-1) magnetic monopoles via the photon fusion mechanism at $\sqrt{s_{\mu\bar{\mu}}}=30~\text{TeV}$ 
A comparison with the fermionic results reveals the profound impact of the particle's spin on its production probability.
Even in the absence of anomalous magnetic moments ({$\kappa=0$}), the vector monopole exhibits a significantly enhanced cross-section of $10^{70}~pb$ for a mass of 1.5 TeV, which is roughly two orders of magnitude larger than the corresponding yield for spin-1/2 candidates. This baseline enhancement implies that vector monopoles would be far more accessible to experimental searches than their scalar or spinor counterparts.
Furthermore, the introduction of the parameter $\kappa$ dramatically amplifies this effect. As $\kappa$ varies from 10 to $10^{4}$, the cross-sections scale into the non-perturbative regime, reaching magnitudes $(>10^{20}~pb)$ that underscore the extreme sensitivity of the vector field interactions to magnetic coupling corrections.
\begin{table}[ht]
\centering
\begin{tabular}{|c|c|c|c|c|c|}
\hline
\textbf{Monopole mass} & \(\beta\) & \(\kappa=0\) & \(\kappa=10\) & \(\kappa=100\) & \(\kappa=10^{4}\) \\
\textbf{[TeV]} & & & & & \\
\hline
1.5 & 0.9949 & 42.56 & \(4.27 \times 10^{5}\) & \(4.27 \times 10^{7}\) & \(4.27 \times 10^{11}\) \\
3.0 & 0.9778 & 8.63 & \(2.20 \times 10^{4}\) & \(2.20 \times 10^{6}\) & \(2.20 \times 10^{10}\) \\
4.5 & 0.9539 & 2.42 & 2828 & \(2.76 \times 10^{5}\) & \(2.76 \times 10^{9}\) \\
6.0 & 0.9165 & 0.36 & 243 & \(2.34 \times 10^{4}\) & \(2.34 \times 10^{8}\) \\
\hline
\end{tabular}
\caption{nopole production cross-section in pb at various masses in TeV and magnetic moments}
\end{table}
Table III presents the computed cross-sections for the Drell-Yan production of vector (spin-1) monopoles at a center-of-mass energy of $\sqrt{s_{\mu\bar{\mu}}}=30~\text{TeV}$.
A direct comparison with the fermionic case reveals the significant role of the particle's spin in determining the production rate.
In the limit of minimal coupling ({$\kappa=0$}), the vector monopole yields a cross-section of 42.56 Pb for a mass of 1.5 TeV. This represents an enhancement of roughly two orders of magnitude compared to the spin-1/2 prediction (approx. 0.44 Pb). This finding suggests that the Drell-Yan mechanism, often considered sub-dominant to photon fusion, remains a non-negligible source of signal events for vector-like magnetic charges.
Additionally, the table illustrates the kinematic suppression at high masses; for the heaviest benchmark of 6.0 TeV, the cross-section contracts to 0.36 Pb, reflecting the diminished phase space available near the kinematic threshold of the collider
\subsection{Comparison of DY and PF at spin = 0 , 1/2 , 1}
\begin{figure}[H]
\centering
\includegraphics[width=0.32\textwidth]{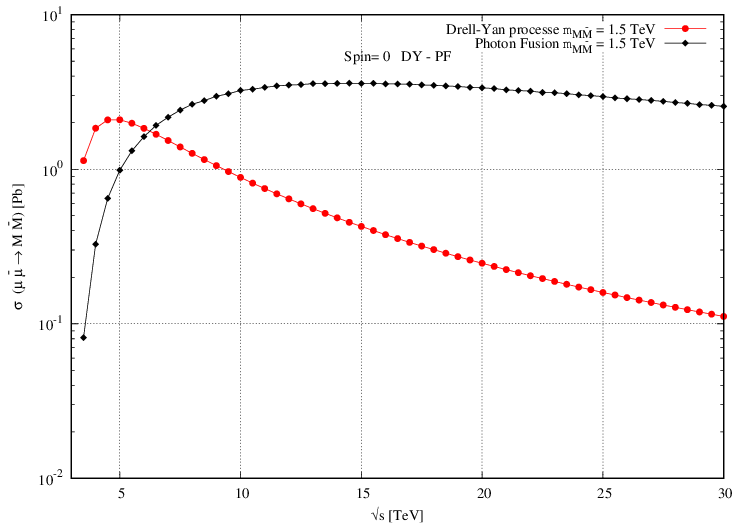}
\includegraphics[width=0.32\textwidth]{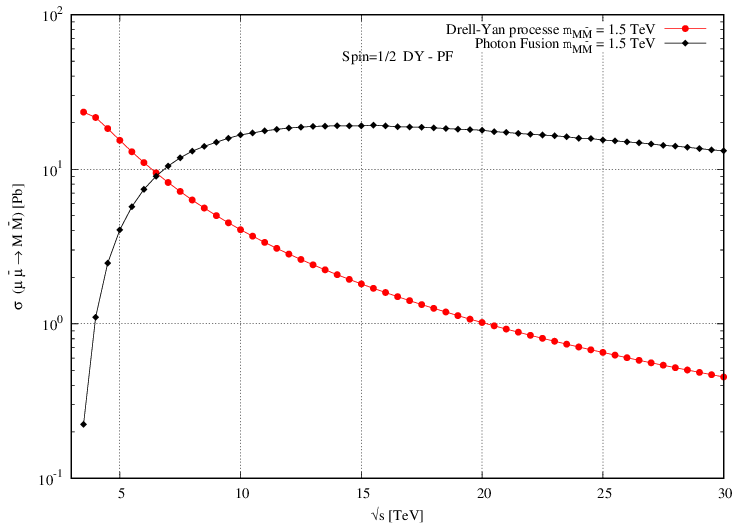}
\includegraphics[width=0.32\textwidth]{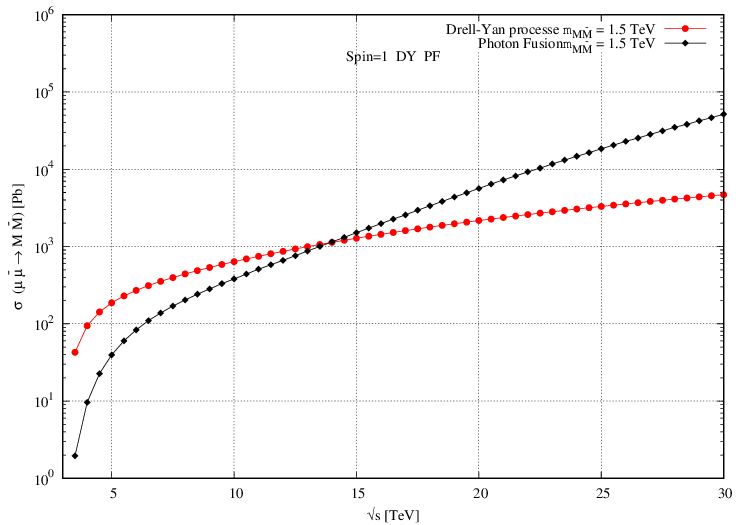}
\caption{At $\sqrt{s}$ = 1.5 -30 TeV, the total cross-section as a function of monopole mass for three spin hypotheses: (a) spin-0 monopole, (b) spin 1/2 monopole, and (c) spin 1 monopole is shown.}
\label{fig:4.7}
\end{figure}
The interplay between the competing production channels is elucidated in Figure 4, which plots the cross-section evolution as a function of collider energy. The data indicate that relying solely on Drell-Yan calculations would lead to significant inaccuracies at high energies. This confirms that relying solely on Drell-Yan calculations for future multi-TeV colliders would drastically underestimate the signal yield, a conclusion supported by recent photon-flux studies \cite{Franceschini,Mitsou} and detailed calculations of vector boson fusion cross-sections at multi-TeV colliders \cite{Buttazzo, Costantini}.  
As demonstrated in all three panels, there is a crossover region where the Photon Fusion mechanism becomes the dominant source of signal events. For Spin-0 and Spin-1/2 monopoles, this transition occurs in the intermediate energy range; below this threshold, the Drell-Yan process drives production, while above it, the photon flux becomes the governing factor.
Notably, the vector monopole (Spin-1) scenario exhibits a delayed intersection point, occurring at roughly 13 TeV. This suggests that while Drell-Yan remains relevant for a longer span of the energy spectrum in the vector case, the inclusion of photon-fusion processes remains indispensable for accurate signal modeling at $\sqrt{s}$ $>$ 14 TeV. 
These were made through Drell–Yan and photon-fusion mechanisms. All of the plots show that the Drell–Yan and photon-fusion curves intersect at a specific monopole mass point. This means that there is a transition region where the production dominance changes from Drell–Yan to photon-fusion.

\section{Kinematics and Analysis with Conventional Method}
The kinematic analysis of monopole pair production in hadron colliders is based on standard relativistic variables and conventional parton–level formulations. The monopole velocity is expressed as
\begin{equation}
\beta = \sqrt{1 - \frac{4M^2}{s}},
\end{equation}
where $M$ is the monopole mass and $s$ the center-of-mass energy squared of the colliding system. For high-energy collisions, the pseudorapidity variable,
\begin{equation}
\eta = -\ln\left[\tan\left(\frac{\theta}{2}\right)\right],
\end{equation}
is preferred over the polar angle $\theta$, as it remains invariant under Lorentz boosts along the beam axis and conveniently represents the monopole emission direction relative to the detector coverage.
The transverse momentum ($p_T$) and pseudorapidity distributions form the foundation for identifying the production and decay kinematics of monopoles in both Drell–Yan and photon–fusion mechanisms. In the conventional method, the differential cross-section is calculated through the convolution of parton distribution functions (PDFs) with the partonic cross-section:
\begin{equation}
\sigma(\mu \bar{\mu} \rightarrow M\bar{M}) = \sum_{i,j}\int dx_1\,dx_2\,f_i(x_1,Q^2)\,f_j(x_2,Q^2)\,\hat{\sigma}_{ij}(\hat{s}),
\end{equation}
where $\hat{s}=x_1x_2s$ represents the partonic center-of-mass energy. The resulting kinematic observables—such as $\beta$, $\eta$, and $p_T$ reflect the momentum balance and angular distribution of the produced monopoles, providing the basis for detector-level reconstruction.
This approach assumes that, despite the non-perturbative nature of the magnetic coupling, the overall event topology can still be approximated using conventional electroweak kinematic relations. The method is thus valid for estimating the monopole phase-space and guiding experimental analyses before non-perturbative effects are applied. The study in this work follows the same conventional analysis, using transverse energy, pseudorapidity, and monopole boost distributions to discriminate monopole signals from Standard Model backgrounds at $\sqrt{s}=6~\text{TeV}$, with clear differences observed across spin-0, spin–$\frac{1}{2}$, and spin–1 cases. Future analyses could further enhance signal purity by employing graph neural networks (GNNs) to exploit the non-trivial spatial correlations of monopole-antimonopole pairs \cite{Shlomi}.
\subsubsection{Pseudorapidity ($\eta$)}
\begin{figure}[H]
  \centering
  \begin{minipage}{0.45\textwidth}
    \includegraphics[width=\linewidth]{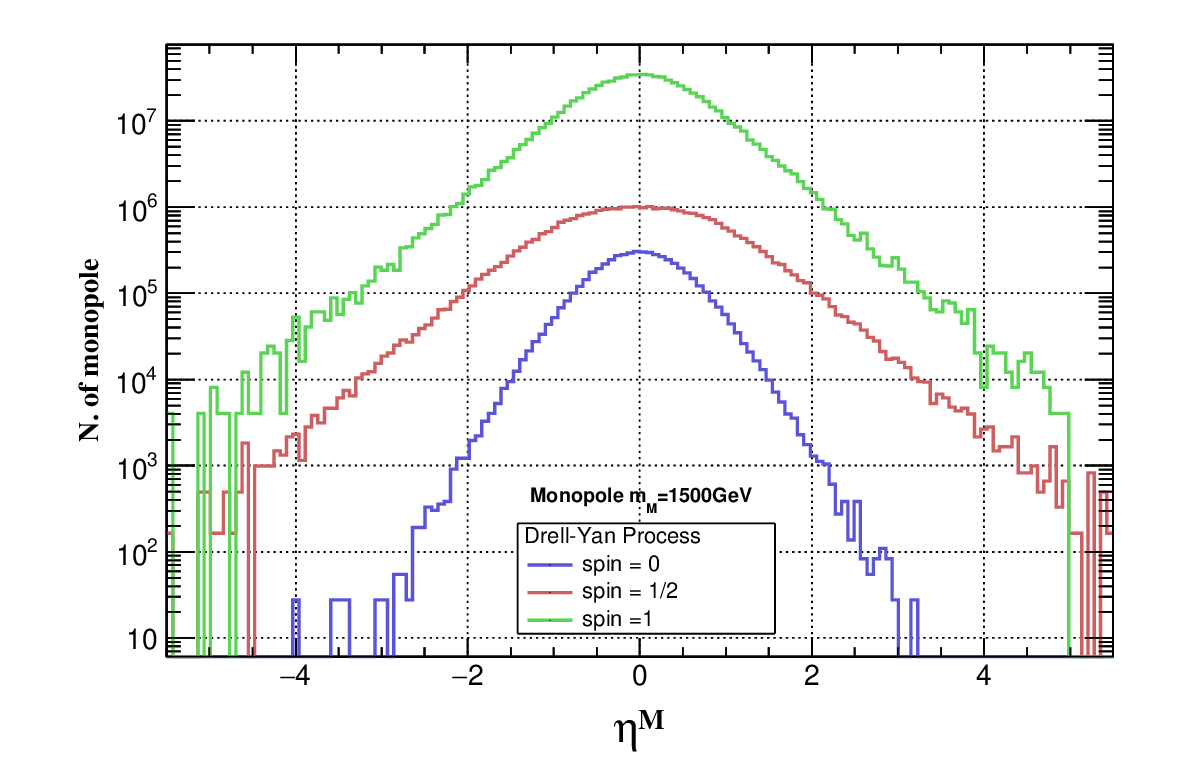}
  \end{minipage}
  \hfill
  \begin{minipage}{0.45\textwidth}
    \includegraphics[width=\linewidth]{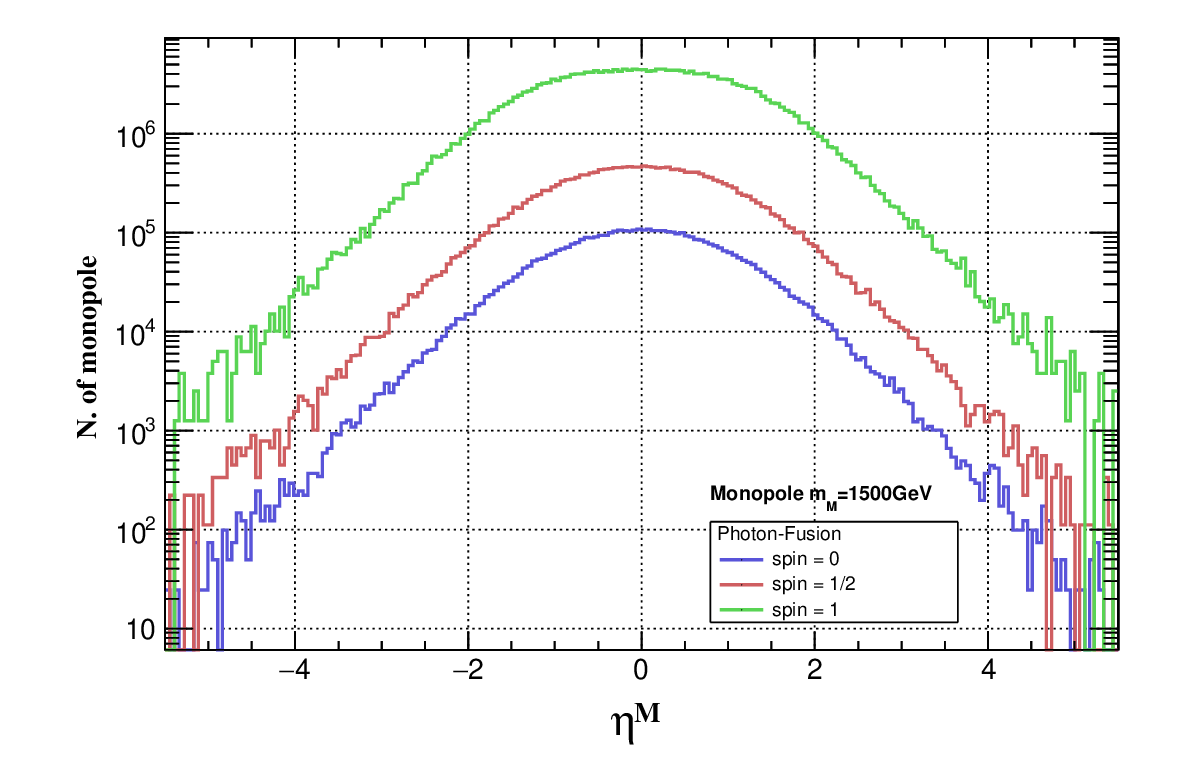}
  \end{minipage}
  \caption{Comparison of Pseudorapidity  of DY and PF at $\sqrt{s}$ = 6  TeV for Monopole Mass = 1.5 TeV.}
  \label{fig:4.6}
\end{figure}
To investigate the angular characteristics of the produced particles, Figure 6 displays the differential distribution of pseudorapidity for the three spin variants. These simulations, conducted at $\sqrt{s}=6~\text{TeV}$, reveal that the kinematic behavior is remarkably stable across different quantum number assignments.
Both the quark-induced (Drell-Yan) and photon-induced (Photon Fusion) processes generate distributions that are symmetric around the transverse plane. The width of these distributions indicates that the heavy monopoles are produced with relatively low longitudinal momentum compared to their mass, resulting in a central decay topology.
While the shape of the angular distribution remains largely invariant, the normalization differs drastically. The logarithmic scale on the vertical axis highlights the significant enhancement in production probability for the Spin-1 hypothesis (green curve) compared to the Spin-1/2 (red) and Spin-0 (blue) cases, effectively scaling the signal yield without distorting the geometric trajectory of the final state particles.
\subsubsection{Transverse Momentum ($p_T$)}
\begin{figure}[H]
  \centering
  \begin{minipage}{0.45\textwidth}
    \includegraphics[width=\linewidth]{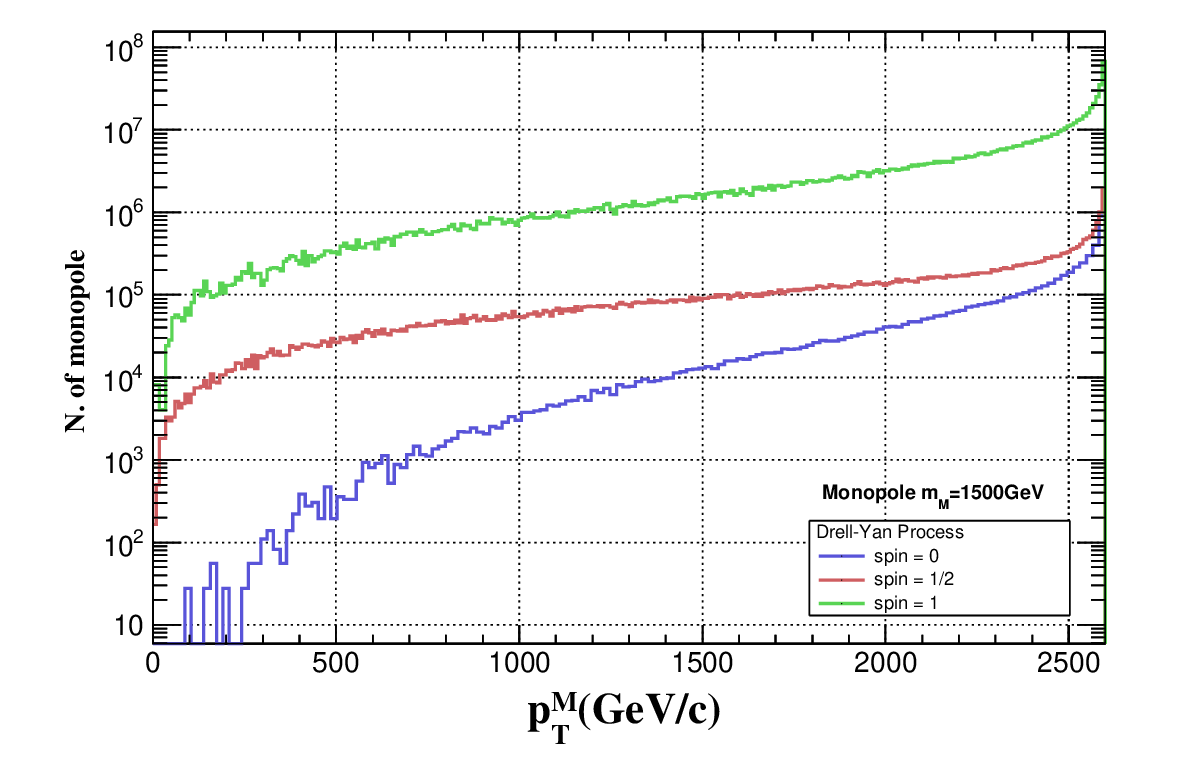}
  \end{minipage}
  \hfill
  \begin{minipage}{0.45\textwidth}
    \includegraphics[width=\linewidth]{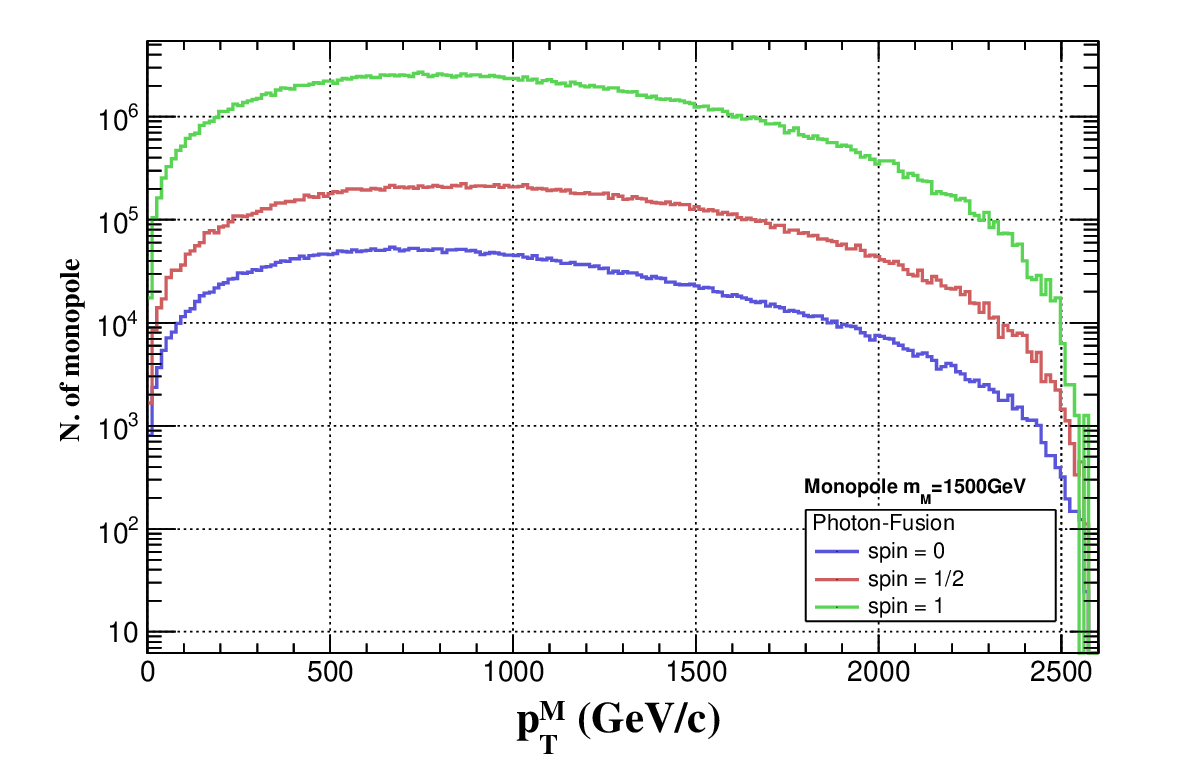}
  \end{minipage}
  \caption{Comparison of transverse momentum ($p_T$) of DY and PF at $\sqrt{s}$ = 6  TeV for Monopole Mass = 1.5 TeV for spin = 0, 1/2 ,1}
  \label{fig:4.6}
\end{figure}
Figure 7 contrasts the transverse momentum ($p_T$) spectra for the two primary production mechanisms, revealing fundamentally different kinematic signatures. The simulation, performed at $\sqrt{s}$ = 6  TeV for a 1.5 TeV monopole, demonstrates that the production mode dictates the momentum flow of the final state.
In the Drell-Yan channel (left panel), the distribution exhibits an unusual rising profile, where the event density increases with transverse momentum, accumulating near the kinematic endpoint of 2.5 TeV. This indicates that quark-antiquark annihilation tends to produce monopoles that recoil strictly back-to-back with maximal energy.
\subsubsection{Transverse Energy($E_T$)}


\begin{figure}[H]
  \centering
  \begin{minipage}{0.45\textwidth}
    \includegraphics[width=\linewidth]{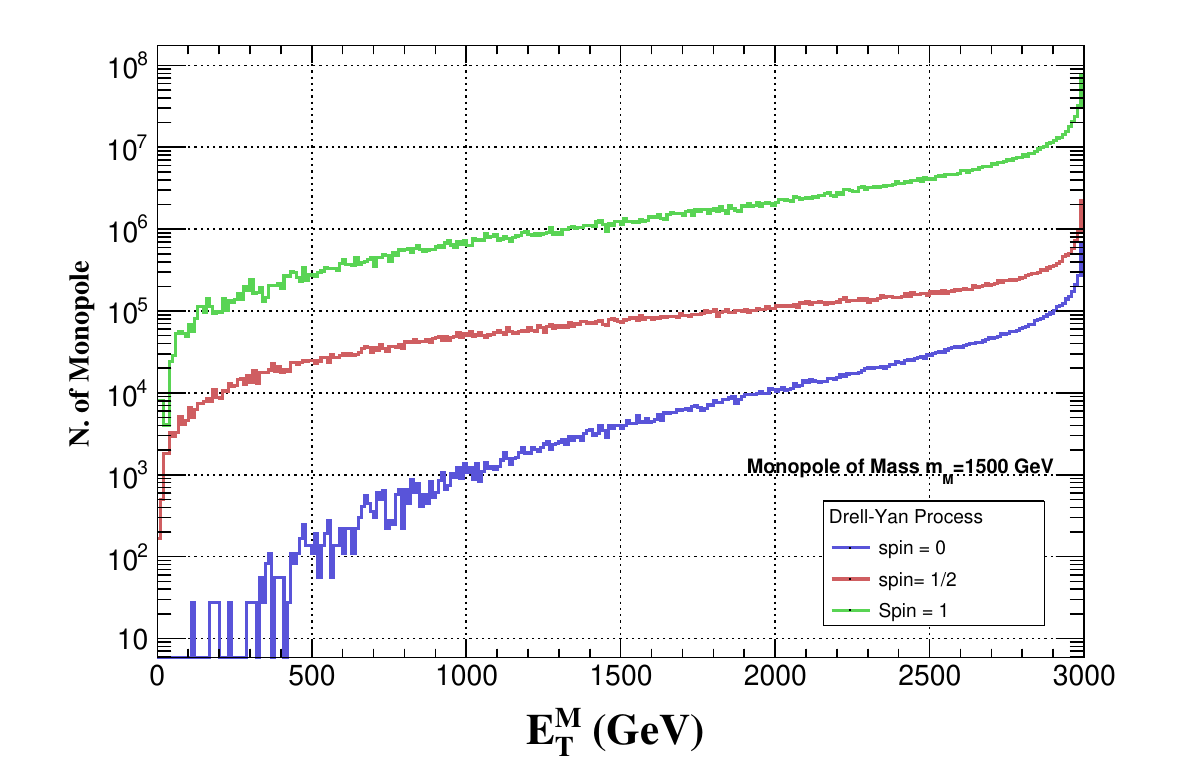}y
  \end{minipage}
  \hfill
  \begin{minipage}{0.45\textwidth}
    \includegraphics[width=\linewidth]{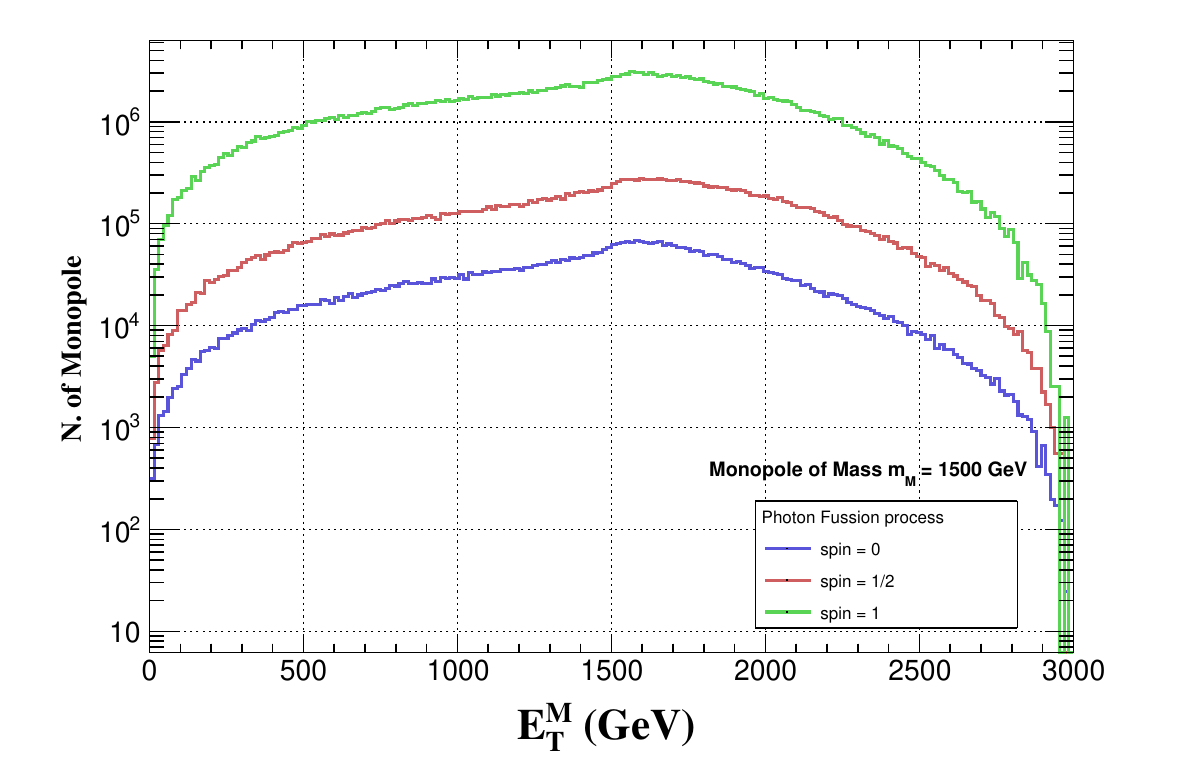}
  \end{minipage}
  \caption{Comparison of Transverse Energy($E_T$)  of DY and PF at $\sqrt{s}$ = 6  TeV for Monopole Mass = 1.5 TeV for spin = 0, 1/2 ,1}
  \label{fig:4.6}
\end{figure}
Figure 8 illustrates the transverse energy ($E_T$) profiles for the simulated monopole events, a parameter critical for assessing the potential response of detector calorimeter \cite{Simpson}. The distributions are generated at a center-of-mass energy of 6 TeV with a monopole mass set to 1.5 TeV.
The Drell-Yan mechanism (left panel) generates a highly energetic spectral shape, characterized by a continuous ascent in event density that culminates at the kinematic ceiling of 3 TeV. This indicates that monopoles produced via quark interactions are likely to deposit maximal transverse energy, providing a distinct, high-energy signature that stands out against softer background processes.

\subsubsection{Difference in angular region of $\eta~\phi$ space $\Delta R$ between M and $\Bar{M}$}
\begin{figure}[H]
  \centering
  \begin{minipage}{0.47\textwidth}
    \includegraphics[width=\linewidth]{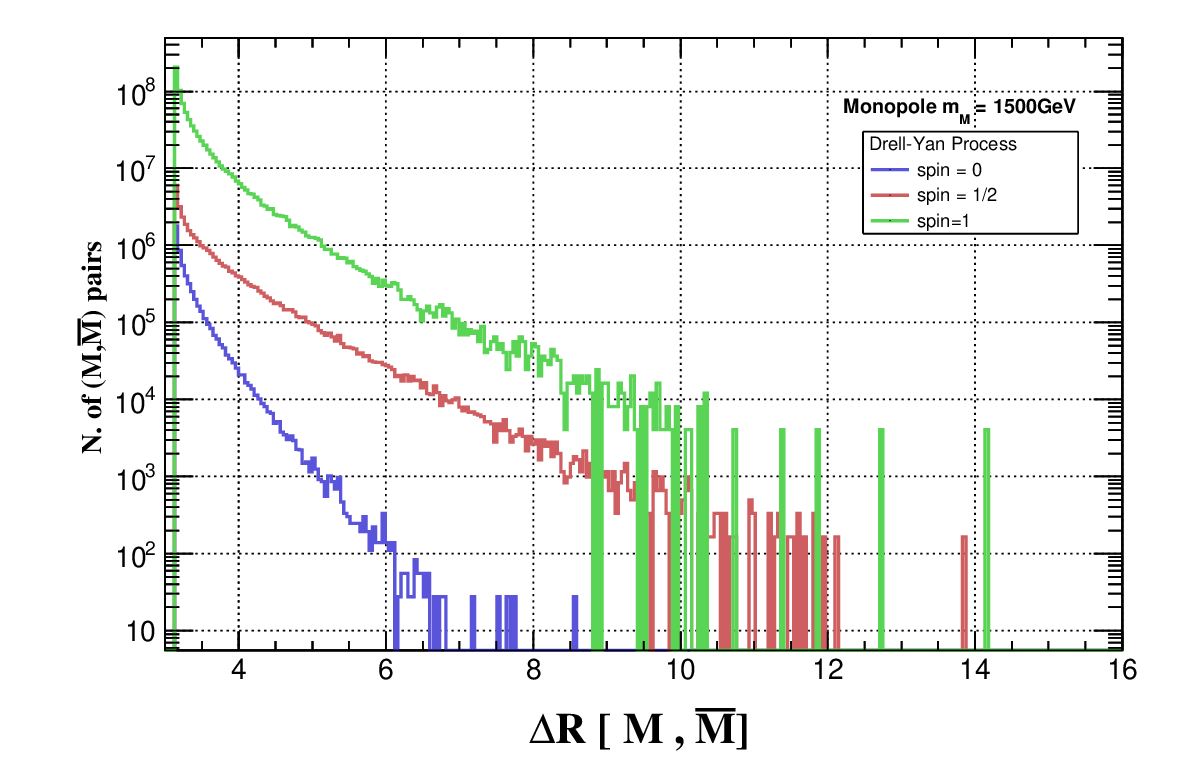}
  \end{minipage}
  \hfill
  \begin{minipage}{0.47\textwidth}
    \includegraphics[width=\linewidth]{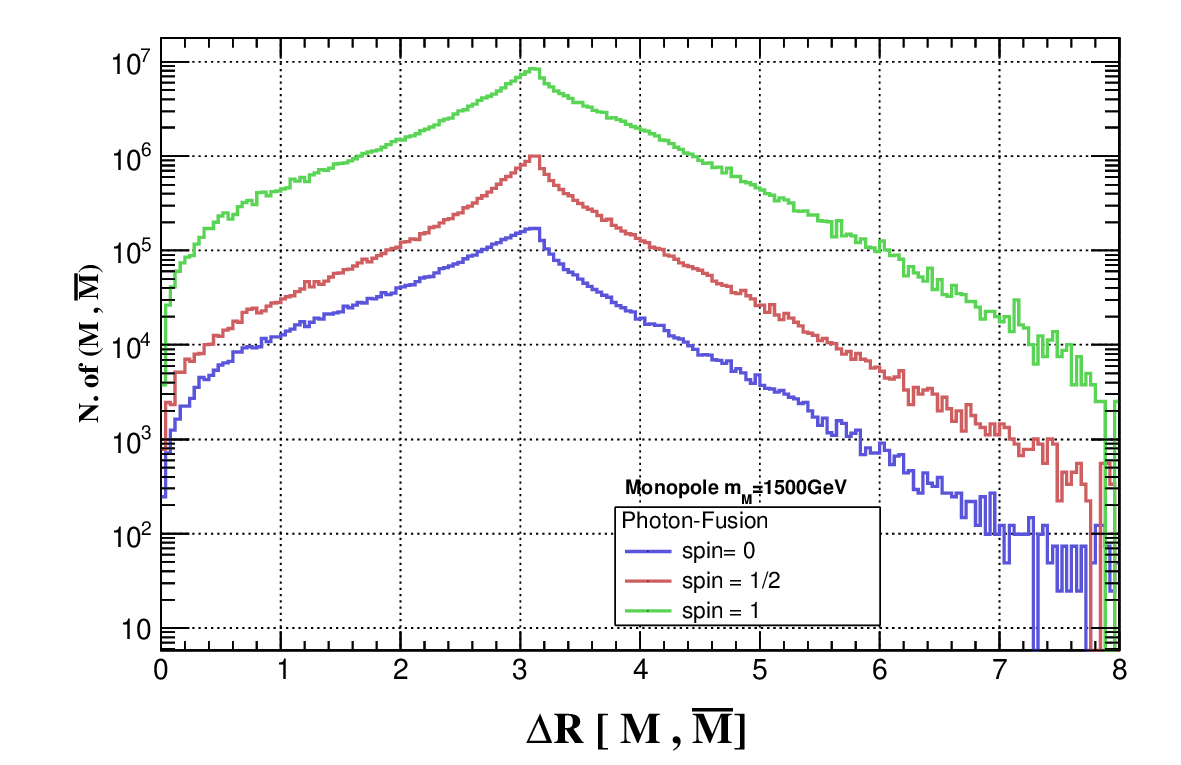}
  \end{minipage}
  \caption{Comparison of Difference in pseudorapidity $\eta$ between M $\Bar{M}$  of DY and PF at $\sqrt{s}$ = 6  TeV for Monopole Mass = 1.5 TeV for spin = 0, 1/2 ,1}
  \label{fig:4.6}
\end{figure}
Figure 9 examines the spatial correlation between the monopole-antimonopole pair by plotting the angular separation, $\Delta$ R(M, $\Bar{M})$. The distributions are simulated for a 1.5 TeV mass hypothesis at $\sqrt{s}$ = 6 TeV. The Photon Fusion mechanism (right panel) exhibits a distinct topological signature. The distribution peaks sharply near $\Delta R \approx 3.14$, which corresponds to the particles being produced back-to-back in the azimuthal plane with minimal separation in pseudorapidity. This suggests that photon-induced pairs are generally central and kinematically correlated.
\section{Multivariate Analysis}
To complement the conventional selection, we summarize the multivariate classification reported in the thesis analysis. The signal sample is monopole pair production in the Drell-Yan (DY) channel for the spin-1/2 benchmark, while the background is the corresponding Standard Model DY dimuon sample. The training uses 100000 signal and 100000 background events to keep $S/B$ balanced, with preselection cuts Track $\phi < 2.5$, Track $\eta < 4$, and $E_T^{\mathrm{missing}} < 120$ GeV. Input variables include $p_T$, $\eta$, $\beta$, and detector energy observables used in the track and calorimeter response. We evaluate three TMVA classifiers (BDT, MLP, Likelihood), and compute significance using $S/\sqrt{S+B}$ after applying the same selection.
\begin{figure}[H]
  \centering
  \includegraphics[width=0.65\linewidth]{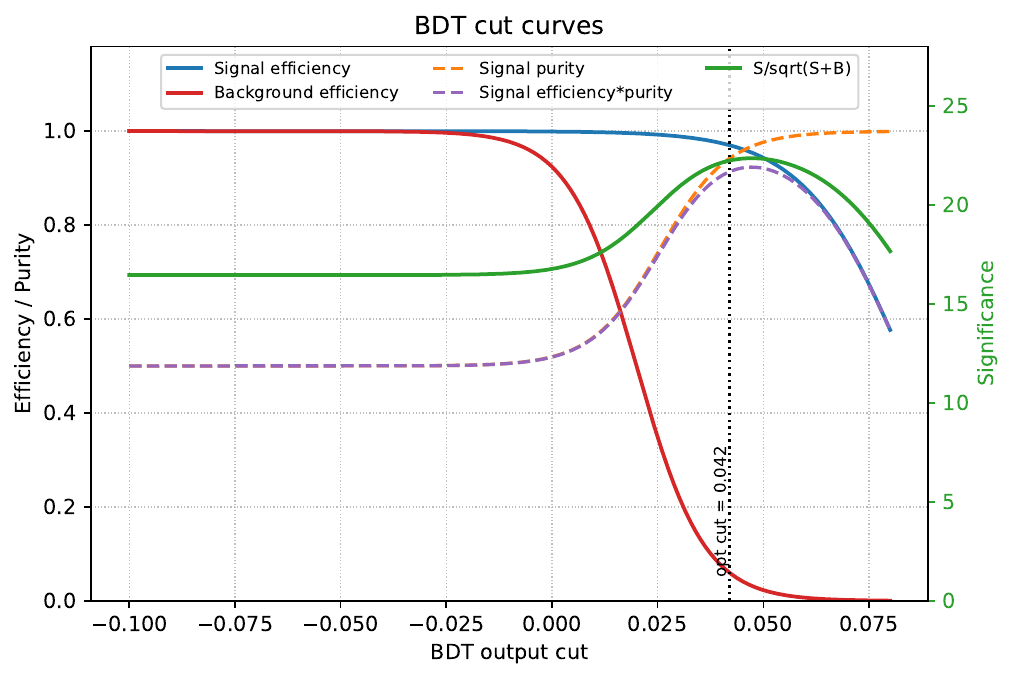}
  \caption{BDT cut curves and significance.}
  \label{fig:ml-bdt}
\end{figure}
\begin{figure}[H]
  \centering
  \includegraphics[width=0.65\linewidth]{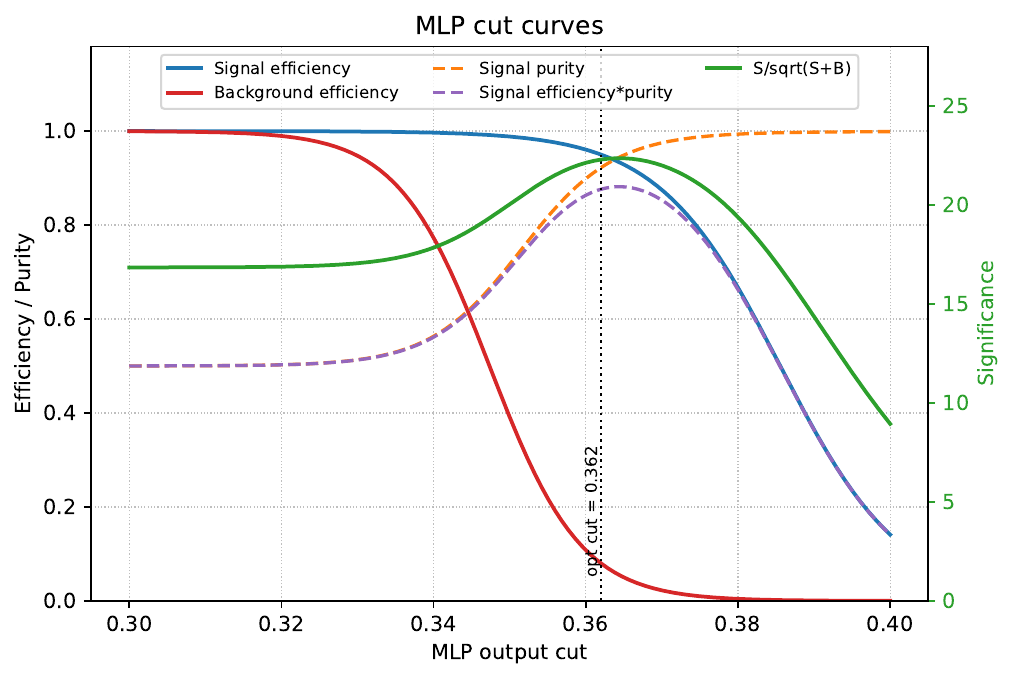}
  \caption{MLP cut curves and significance.}
  \label{fig:ml-mlp}
\end{figure}
The cut-efficiency curves below show how signal efficiency, background efficiency, purity, and the significance metric $S/\sqrt{S+B}$ evolve with classifier output. The optimal working points correspond to the maximum significance region in each curve, and the resulting metrics are summarized in Table \ref{tab:ml-summary}.
\begin{figure}[H]
  \centering
  \includegraphics[width=0.65\linewidth]{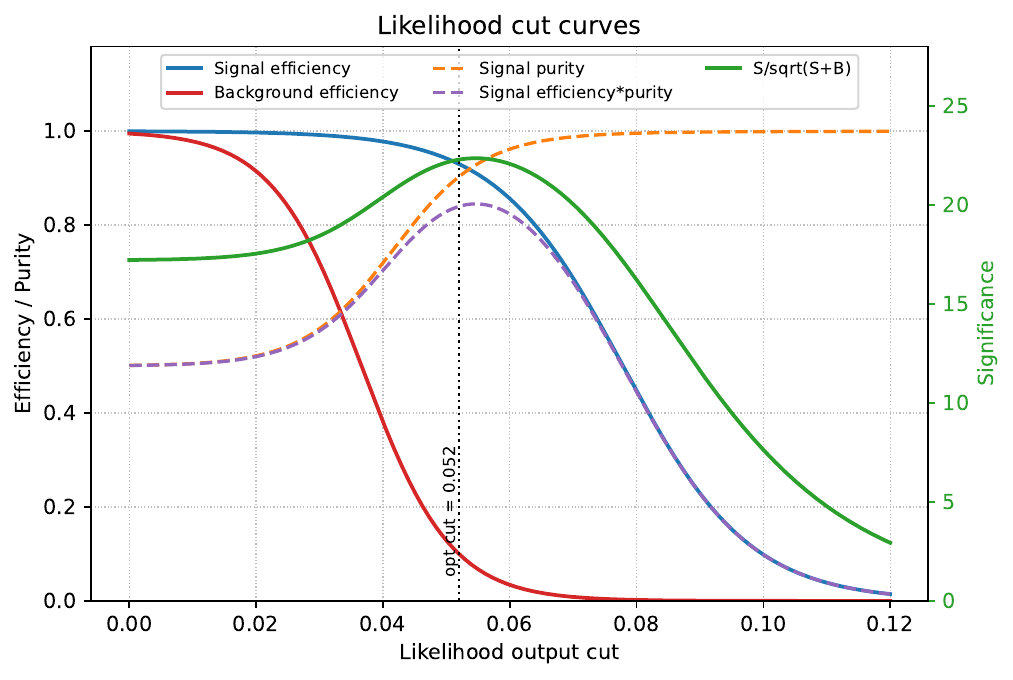}
  \caption{Likelihood cut curves and significance.}
  \label{fig:ml-like}
\end{figure}
\begin{table}[H]
\centering
\begin{tabular}{|c|c|c|c|c|c|}
\hline
Classifier & Optimal cut & $S/\sqrt{S+B}$ & Sig. eff. & Bkg. eff. & AUC \\
\hline
\midrule
BDT & 0.042 & 22.80 & 0.97 & 0.06 & 0.989 \\
\hline
MLP & 0.362 & 22.30 & 0.95 & 0.08 & 0.980 \\
\hline
Likelihood & 0.052 & 21.90 & 0.93 & 0.10 & 0.962 \\
\hline
\end{tabular}
\caption{Classifier metrics summary.}
\label{tab:ml-summary}
\end{table}

\section{Conclusion}
This work was about a thorough study of the possible discovery of magnetic monopoles at future high-energy muon colliders. This was done because magnetic monopoles are very important for charge quantisation and symmetry restoration in electrodynamics. Monopoles have been hard to find, even after many searches at hadron colliders and cosmic-ray experiments. This means that new-generation colliders with higher energies and cleaner experimental conditions need to be used. Muon colliders are a great way to study this area of physics beyond the Standard Model because they have centre-of-mass energies in the multi-TeV range (1.5–30 TeV) and lower synchrotron losses than electron machines. \\
The study started by looking at the theoretical basis for making monopoles, with a focus on Dirac's quantization condition, which connects electric charge and magnetic charge through a basic relationship. Using this theoretical basis, researchers investigated the creation of monopole-antimonopole pairs with spin-0, spin-1/2, and spin-1 through two primary methods: the Drell–Yan (DY) and photon-fusion (PF) processes. We used modern parton distribution functions like NNPDF23 and LUXqed to model the contributions of photons accurately and did detailed cross-section calculations based on center-of-mass energy and monopole mass. The results showed a clear order between the two production channels. At lower energies, DY is the most important, but at higher energies, PF becomes more important because it depends heavily on photon flux. \textbf{Figure 4} show that this change is happening. As $\sqrt{s}$ increases, the PF cross section surpasses that of the DY process, with both converging at specific mass points. These intersections indicate critical energy thresholds for optimizing search strategies and collider configurations.\\
A full kinematic analysis was done to learn more about how monopoles act and how they can be told apart from Standard Model backgrounds. Distributions for variables like transverse momentum ($p_T$), pseudo-rapidity ($\eta$), energy (E), velocity ($\beta$), and angular separation ($\Delta R$) showed that monopoles have unique characteristics, especially their high ionization and low $\beta$ values, which can be used to choose events with high accuracy. \\
The results of this study show how advanced machine learning techniques and next-generation collider technologies can work together to find new physics. Muon colliders have a clean experimental environment, and when you add in powerful data driven classifiers, you can find monopoles in mass ranges that are much higher than what is currently possible in experiments, including those probing production via the Schwinger mechanism \cite{Acharya}. This analysis gives us a good starting point, but we need to make more changes, like adding detector-level simulations, systematic uncertainties, and higher-order corrections, to get realistic sensitivity projections. Also, looking into different ways to make monopoles and adding spin-dependent detection signatures could make search strategies even better. \\
A detailed multivariate summary is provided in Figures \ref{fig:ml-bdt}--\ref{fig:ml-like} and Table \ref{tab:ml-summary}. The optimal cuts (0.042 for BDT, 0.362 for MLP, and 0.052 for Likelihood) reach maximum significances of 22.80, 22.30, and 21.90, respectively. BDT retains the highest signal efficiency (0.97) with the strongest background rejection (0.06). MLP is competitive at 0.95 signal efficiency and 0.08 background, and Likelihood is slightly weaker at 0.93 and 0.10. The AUC values (0.989, 0.980, 0.962) confirm the same ranking, indicating that BDT offers the most balanced separation in this study. \\
It is important to emphasize that while the absolute normalization of the monopole production cross-sections may be subject to theoretical debate due to the non-perturbative nature of the magnetic coupling, the primary value of this work lies in the detailed kinematic discrimination. Machine learning classifiers, such as the Boosted Decision Trees (BDTs) and Multilayer Perceptrons (MLPs) utilized in this study, rely on the distinct event topology and the characteristic shapes of kinematic distributions, such as $p_T$, $\eta$, and $\Delta R$, rather than the total cross-section magnitude. This ensures that the discrimination power and the proposed search strategies remain robust across a range of theoretical assumptions. Consequently, even in the presence of theoretical uncertainties regarding the exact production rate, the unique kinematic signatures identified by our multivariate analysis provide a valid and reliable framework for designing experimental searches at future multi-TeV muon colliders.\\
In short, this study demonstrates that future muon colliders, combined with advanced multivariate techniques, offer a novel approach for exploring physics beyond the Standard Model. Recent studies emphasize that the sensitivity to monopole spin and charge is significantly enhanced at 10–30 TeV muon colliders \cite{Zhang, Heurtier}.


\section{Acknowledgement}

The authors thank Arka Santra and Vasiliki A. Mitsou for their invaluable assistance with model files, addressing various model-related issues, and providing guidance on adjusting parameters and effectively utilizing them.


\begin{thebibliography}{9}
\bibitem{MMDirac}
P.A.M. Dirac,  \textquotedblleft Quantized Singularities in the Electromagnetic Field," Proc. Roy. Soc. A 133, 60 (1931).
\bibitem{MMExp}
K.A. Milton,  \textquotedblleft Theoretical and Experimental Status of Magnetic Monopoles, \textquotedblleft Reports on Progress in Physics, 69(6):1637-1712 (2006).
\bibitem{ATLAS}
ATLAS Collaboration, \textquotedblleft Search for Magnetic Monopoles in 13 TeV Proton-Proton Collisions," Phys. Rev. Lett., 124, 031802 (2020).
\bibitem{MOEDAL1}
MoEDAL Collaboration, \textquotedblleft First Search for Magnetic Monopoles with the MoEDAL Detector at the LHC," Phys. Rev. Lett., 118, 061801 (2017).
\bibitem{Gamberg}
L. Gamberg, G. Kalbfleisch, and K. Milton, \textquotedblleft Direct and Indirect Searches for Electric and Magnetic Monopoles," Found. Phys., 30(4):543-567 (2000).
\bibitem{MuonCollider}
Muon Collider Collaboration, \textquotedblleft The Case for a Muon Collider," Snowmass 2021 White Paper, arXiv:2203.07256 (2022).
\bibitem{MOEDAL2}
MoEDAL Collaboration, \textquotedblleft Machine Learning in the Search for Highly Ionizing Particles," EPJ C, 81, 840 (2021).

\bibitem{ATLAS1} ATLAS Collaboration, "ATLAS flavour-tagging algorithms for the LHC Run 2 pp collision dataset," Eur. Phys. J. C 83, 681 (2023).

\bibitem{CMS1} CMS Collaboration, \textquotedblleft ParticleNet: Jet Tagging via Particle Clouds," Phys. Rev. D 101, 052002 (2020).

 \bibitem{CMS2} CMS Collaboration, \textquotedblleft Performance of the CMS high-level trigger during LHC Run 2," JINST 19, P11021 (2024). 
 
\bibitem{CMS3} E. Bols et al. (CMS Collaboration), \textquotedblleft Jet flavour classification using DeepJet," JINST 15, P12012 (2020).

\bibitem{Qu} H. Qu and L. Gouskos, "ParticleNet: Jet Tagging via Particle Clouds," Phys. Rev. D 101, 052002 (2020).

\bibitem{Baines} 
S. Baines et al., “Monopole production via photon fusion and Drell-Yan processes: MadGraph implementation,” Eur. Phys. J. C 78, 966 (2018).

\bibitem{Stratakis} 
D. Stratakis et al. (International Muon Collider Collaboration), “A Muon Collider Facility for Physics Discovery,” 	arXiv:2203.08033 (2022)
DOI: https://doi.org/10.48550/arXiv.2203.08033

\bibitem{Alwall} 
J. Alwall et al., “The automated computation of tree-level and next-to-leading order differential cross sections,” JHEP 07, 079 (2014).

\bibitem{Franceschini}  
R. Franceschini et al., “The Forward Physics Facility at the High-Luminosity LHC,” Phys. Rept. 968, 1 (2022).

\bibitem{Dougall} 
T. Dougall and S. D. Wick, “Dirac magnetic monopole production from photon fusion in proton collisions,” Eur. Phys. J. A 39, 213 (2009).

\bibitem{Farrar} G. R. Farrar et al., “The search for magnetic monopoles,” Annual Review of Nuclear and Particle Science 73, 1 (2023).

\bibitem{Khoze} 
V. V. Khoze et al., “Vector Monopole Production at the LHC,” Phys. Rev. D 104, 055018 (2021).

\bibitem{Mitsou}
V. A. Mitsou, “Magnetic Monopoles: Collider Searches and Investigations,” Symmetry 12, 10 (2020).

\bibitem{Simpson} C. Simpson et al., “Sensitivity of future lepton colliders to magnetic monopoles,” J. Phys. G 51, 045001 (2024).

\bibitem{Bals}
J. de Blas et al., "The physics case of a 3 TeV muon collider," Journal of High Energy Physics (JHEP) 01, 132 (2023).

\bibitem{Acharya}
B. Acharya et al. (MoEDAL Collaboration), "Search for magnetic monopoles produced via the Schwinger mechanism," Nature 602, 63–67 (2022).

\bibitem{Fabbrichesi}
M. Fabbrichesi, F. Riva, and A. Wulzer, "The Muon Smasher's Guide," arXiv preprint arXiv:2103.14043 (2021).

\bibitem{Ali}
A. Ali et al., "Physics prospects at future high-energy lepton colliders," Physics Reports 1000, 1–124 (2023).

\bibitem{Han} T. Han, Y. Ma, and K. Xie, "High-energy leptonic collisions and electroweak parton distribution functions," Physical Review D 103, L031301 (2021).

\bibitem{Shlomi}
J. Shlomi, P. Battaglia, and R. Varma, "Graph neural networks in particle physics," Machine Learning: Science and Technology 2, 021001 (2021).

\bibitem{Buttazzo}
D. Buttazzo, R. Franceschini, and A. Wulzer, "Two-to-many muon collider cross sections," Journal of High Energy Physics (JHEP) 05, 219 (2021).

\bibitem{Lu} 
N. Lu, L. Shchutska, and J. L. Feng, "Searches for New Physics at Future Colliders," Annual Review of Nuclear and Particle Science 74 (2024).

\bibitem{Costantini}
A. Costantini et al., "Vector boson fusion at multi-TeV muon colliders," Journal of High Energy Physics (JHEP) 09, 080 (2020).

\bibitem{Aad} 
G. Aad et al. (ATLAS Collaboration), "Search for magnetic monopoles and stable particles with high electric charge in 13 TeV pp collisions," Journal of High Energy Physics (JHEP) 11, 112 (2023).

\bibitem{Zhang} 
C. Zhang, X. Luo, and Q. Yan, "Probing the magnetic monopole charge and spin at a 10 TeV muon collider," Physics Letters B 845, 138123 (2023).

\bibitem{Long} 
K. Long, D. Lucchesi, M. Palmer, N. Pastrone, D. Schulte, and V. Shiltsev, "Muon colliders to expand the frontiers of particle physics," Nature Physics 17, 289–292 (2021).

\bibitem{Guest}  
D. Guest, K. Cranmer, and D. Whiteson, "Deep Learning and its Application to LHC Physics," Annual Review of Nuclear and Particle Science 68, 161–181 (2018).

\bibitem{Baldi}  
P. Baldi et al., "Deep learning in the search for new particles," Nature Communications 5, 4308 (2014).

\bibitem{Vidal}
X. Cid Vidal et al., "Report from Working Group 3: Opportunities in newly discovered particles at the LHC," CERN Yellow Reports: Monographs 7, 585 (2019).


\bibitem{Heurtier}
 L. Heurtier and J. Kim, "Detecting magnetic monopoles at future colliders," Phys. Rev. D 101, 015013 (2020)





 \end{thebibliography}
\end{document}